# Au-assisted substrate-faceting for inclined nanowire growth

*Jung-Hyun Kang[†][*], Filip Krizek[‡][*], Magdalena Zaluska-Kotur[§], Peter Krogstrup[‡], Perla Kacman[§], Haim Beidenkopf[†] and Hadas Shtrikman[†]*

† Dept. of Condensed Matter Physics, Braun Center for Submicron Research, Weizmann Institute of Science, Rehovot 76100, Israel

‡ Center for Quantum Devices and Station Q Copenhagen, Niels Bohr Institute, University of Copenhagen, 2100 Copenhagen, Denmark

§ Institute of Physics Polish Academy of Science, Al. Lotnikow 32/46, 02-668 Warsaw, Poland

ABSTRACT: We study the role of Au droplets in the initial stage of nanowire growth via the vapor-liquid-solid method. Apart from serving as a collections centers for growth species, the Au droplets carry an additional crucial role that necessarily precedes the nanowire emergence, i.e., they assist the nucleation of nano-craters with strongly faceted {111}B side walls. Only once these facets become sufficiently large and regular, the Au droplets start nucleating and guiding the growth of nanowires. We show that this dual role of the Au droplets can be detected and monitored by high-energy electron diffraction during growth. Moreover, Au-induced formation of craters and the onset of nanowires growth on the {111}B facets inside the craters are confirmed by the results of Monte Carlo simulations. The detailed insight into the growth mechanism of inclined nanowires will help engineering new and complex nanowire based device architectures.

INTRODUCTION: Semiconducting nanowires (NWs) of both, III–V type (including InAs and GaAs) and II-VI type (including ZnTe and CdTe), grow preferentially along the <111> direction, following the dangling bonds of the substrate. This takes place regardless of the orientation of the substrate used.[1-3] The main reason behind this preferential direction is that the free energy of the NWs is the lowest along the <111> direction.[4, 5] As a consequence, semiconductor NWs are typically grown on a (111)B (anion terminated) substrate surface, where they emerge vertically.[6-8] Still, during the past decade other low-index-oriented III–V NWs have been sporadically reported.[9-15] These NWs often possess superior structural properties to the <111> oriented NWs such as increased length, the lack of stacking faults, or the support of higher doping levels.[1] Their inclined orientation is also advantageous for various processes, such as robust side coating[16] and high luminescence.[17-19] They also enable formation of intersections and networks,[20-24] important for non-abelian braiding operations,[25, 26] solar cell devices,[27, 28] and biological sensing.[29]

In **Figure 1** the emergence of InAs NWs on four different substrates oriented in all three singular [(001), (011), (111)] crystal planes is compared. First we note that the nucleation of NWs on the (111) and (011) surfaces (**Figures 1a** and **1b**, respectively) starts from extruded nano-piramid structures, while NWs growth on (001) and its related vicinal surface (311)B requires formation of craters with {111}B nano-facets, as seen at **Figures 1c** and **1d**, respectively. In the latter case the Au-assisted nucleation of inclined NWs must be, therefore, preceded by {111}-faceting of the substrate surface, which is also Au-assisted. The main cause for this distinction is the difference in migration length of the impinging atoms over the various surfaces.[30]

As shown in **Figure 1**, on a (111)B surface NWs grow normal to the substrate (**Figure 1a**), while on the (011) they are tilted, all at the same inclining angle between the (011) and (111)B surfaces of about 55° (**Figure 1b**). We note that in both cases the NWs strictly follow the predetermined growth direction appearing parallel to each other. In **Figure 1c** InAs NWs can be seen emerging from a (311)B substrate, as well at a tilt of 55° to the surface. This results from the fact that craters formed on the (311)B surface are strongly affected by the tilt of this surface with respect to the (001) plane, what eliminates one set of {111}B of their side walls. These facets cannot attain a large enough collection area to promote NWs nucleation and consequently NWs can emerge only from the other set of {111}B facets. Thus, again all NWs grow parallel to each other following the same <111> direction. Finally, on the (001) surface, shown in **Figure 1d**, the Au-induced faceting results in formation of craters with two mirror-symmetric opposite {111}B side facets. These are tilted at 35.3° to the (001) substrate surface and lye parallel to the <110> direction. In this case inclined InAs NWs can emerge in two different <111> directions. These results show, therefore, that the formation of NW intersections[20-24, 31] is possible only on <001>-oriented substrates. Thus, while we have studied different surface orientations, as shown in **Figure 1**, and other materials (see Supplementary Information), in the following we concentrate mostly on the growth of the InAs NWs on the (001) InAs surface which are the key players in current mesoscopic physics experiments, which requires NW networks.

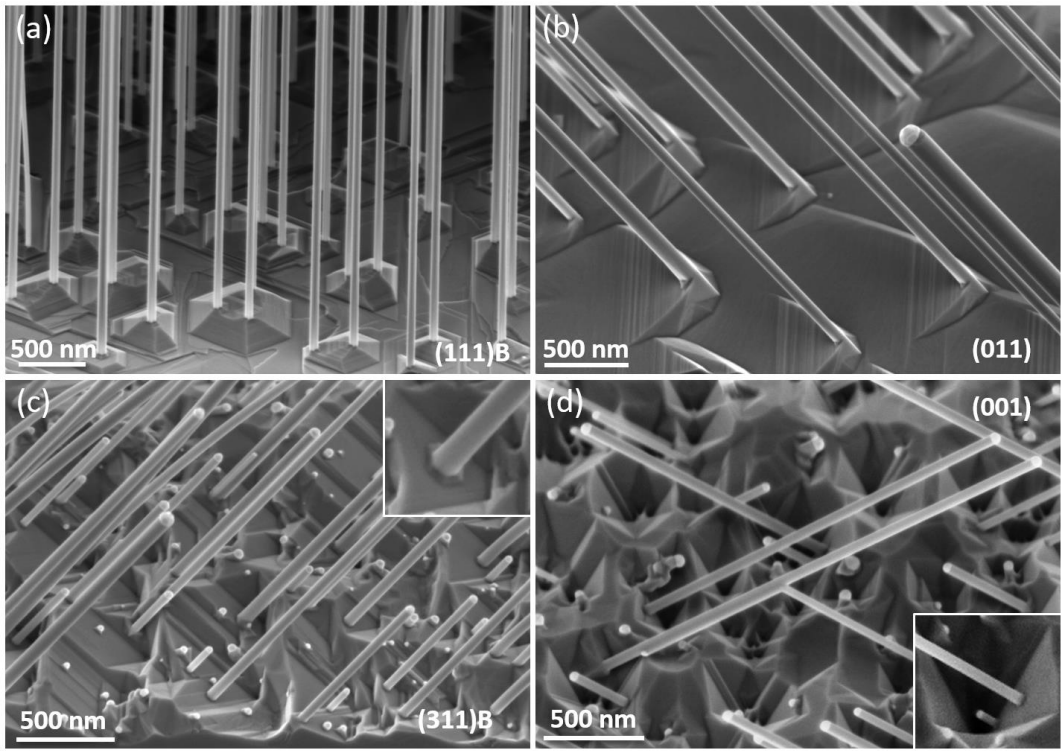

**Figure 1**: **(a)-(d)** SEM images of NWs grown on four different surfaces: (111)B, (110), (311)B and (001), respectively. In **(a)** and **(b)** the NWs are seen to emerge from faceted nano-pyramids whereas in **(c-d)**, i.e., on the (001)-type surfaces, they emerge from craters with {111}B facets (as shown in the insets).

EXPERIMENT: The NWs are grown in a high purity MBE (RIBER-32) system, in which the InAs substrate first undergoes an oxide blow-off process at ~450 °C with no As overpressure, in a dedicated chamber attached to the growth system. Oxide blow-off is followed by cooling down to ~200 °C where Au evaporation takes place. A thin (<1 nm) Au layer is evaporated while the Au is at 1100 °C and the substrate is extremely close (~15 cm). Thus, the radiation from the cell heats up the substrate and assures it is In rich, given only residual As is present in the chamber. The Au evaporation takes place for 100 - 150 secs. Sample is then moved to the growth chamber and held at 350 - 400 °C until As overpressure builds up to $5 \times 10^{-6}$ torr. In order to form Au droplets the substrate is heated to ~550 °C under As overpressure, where the Au droplets formed experience Oswald ripening on one hand and dissolution of the substrate on the other hand. Thus, they sporadically spread over the substrate. At this initial stage the Au droplets etch into the InAs substrate just under them by forming an In-Au eutectic.[32] This nucleates a nano-crater beneath the Au droplets with two {111}B side facets amidst the (001) surface. The Au droplets also diffuse up to the (001) surface and leave behind a trail of exposed nano-craters (the Au droplet dynamics will be discussed below). This entire process takes place under an overpressure of As flux.

Alternatively, the (111)B substrates were selectively patterned by the Au catalysts in a standard electron beam lithography (EBL) process. The resulting Au islands with diameter of ~ 80 nm and thickness ~ 7 nm, were annealed into Au droplets during substrate oxide desorption growth step at ~ 540 °C under As overpressure ($1.5 \times 10^{-5}$ torr) in a GEN II Varian MBE system. The NWs were grown under similar conditions as stated above.

The InAs NWs and craters on the surface of (001) InAs substrates were characterized by field emission scanning electron microscopy (FE-SEM, Zeiss Supra-55, 3 $kV$, working-distance ~4 mm and FE-SEM JEOL 7800F, 15kV, working distance 15 mm), transmission electron microscopy (TEM), high resolution (HR-) TEM (JEOL JEM-2100 TEM, 200 $kV$, Gatan Ultrascan XP 2k × 2k CCD camera), and atomic force microscopy (AFM, Veeco Digital Instrument Dimension D3100, 17 μm of tip height). The structural properties, including atomic configuration and facet planes, were analyzed using CrystalMaker® for Windows (Ver. 9.2.9f1, CrystalMaker Software Ltd.).

We first note, that the typical two-stage growth protocol on an (001) surface is well captured by reflection high-energy electron diffraction (RHEED) snapshots. At the initial state after oxide blow-off and Au deposition a clear RHEED pattern of the (001) surface is seen in **Figure 2a.** As growth commences, the Au droplets nucleate nano-craters and rhombohedral reflection RHEED pattern reflected from the (001) shoulders of the craters platform begins to form, as shown in **Figure 2b**. This crater incubation stage typically lasts for 10 ~ 20 minutes under given growth conditions. During this period, no RHEED signature of NWs growth is detected. It ends once the {111}B side facets of the craters surpass a critical size needed to support In supersaturation in the Au droplets. This marks the onset of the second stage, in which NWs begin to emerge from the craters along the <111>-orientation. This gives rise to the fragmented RHEED pattern in **Figure 2c**.

The nano-craters take various hexagonal shapes with different proportions of the sidewalls and either a sharp or a flat bottom. A top view SEM image of a typical crater is seen in **Figure 2d** and schematically identified in **Figures 2e**. The symmetric structure can be seen in a top view AFM mapping of a single crater in **Figure S2**. A line scan across the centre of the crater indeed exposes its depth and triangular shape. All craters have an elongated structure, which results from the different sticking probability and diffusion length on the {110} and {111} facets constructing them (here, (111), (-1-11) denote the {111}B and (11-1), (1-11) denote the {111}A, respectively). It should be emphasised that all craters on (001) surface are mirror symmetric, regardless of their shape. Namely, the side facets facing each other have the same type of planes. However, the craters on the (311)B surface are distorted with respect to the mirror plane between both {111}B facets **(Figure 1c)** as a result of the anisotropic diffusion over this surface.

We begin by studying the initial growth stage of craters formation. During the formation of catalyst nanoparticles, the substrate is locally dissolved in a reaction with the Au layer and low-energy {111}B facets develop beneath it. Such an etching process was used before to induce NW growth in the <001> direction, normal to the (001) substrate surface by quenching the temperature.[2, 10] Recent calculation further shows that based on Si dissolution rates craters are formed in the (001) not (111)B surface, since (111)B surfaces are denser, the atoms on this surface are more bound and, therefore, harder to dissolve than on the {001} surfaces.[32] We find that if normal growth conditions are kept, the Au etching nucleates the nano-craters that we regularly find on the (001) surface. While the impinging In atoms diffuse slowly over the corrugated (001) surface, they exhibit high diffusivity over the previously nucleated {111}B nano-facets. Therefore, we speculate that the In atoms have much higher probability to bind to the (001) surface than to the {111}B side facets of the nucleated craters. Consequently, as the 2D growth on the (001) surface progresses, the In atoms keep escaping from the craters, which results in deepening and extending the area of the facets they are comprised of.

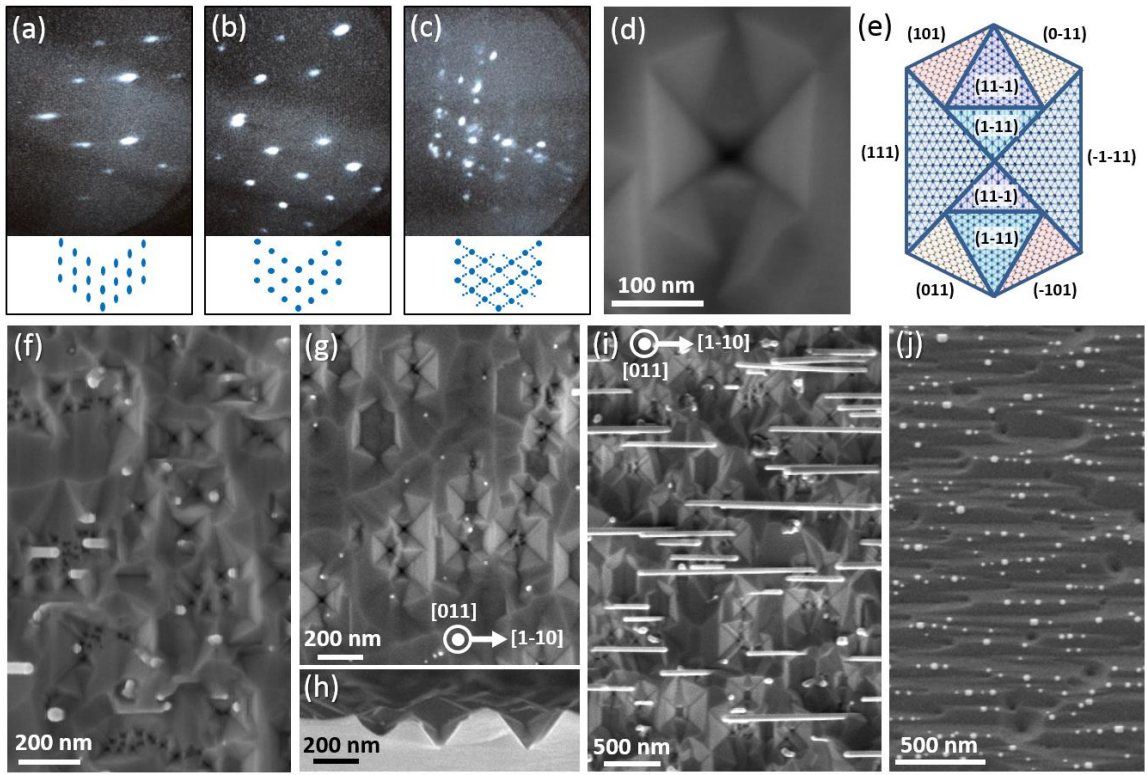

**Figure 2. (a-c)** RHEED patterns: **(a)** the (001) surface Au coated after oxide blow-off; **(b)** craters formation; **(c)** NWs growth. **(d)** Top view of a typical InAs crater. **(e)** Schematic illustration of the crater with relevant facets. **(f-j)** SEM images: **(f)** Top view of the initial stage (10 min) of NWs growth on (001) surface showing the immediate nucleation of craters. **(g)** Top and **(h)** side view of well-defined craters after a prolonged growth (here gold droplets were too small to compete with the bulk growth). **(i)** Top view of InAs NWs on an (001) substrate emerging in two opposite {111}B along the <110> direction. Craters are observed in the background. **(j)** (001) InAs surface passivated under As overpressure, showing no craters (and no NWs) have formed by the gold droplets.

The nano-crater evolution is captured by SEM in **Figures 2f-i**. A (001) substrate where growth was deliberately halted during the crater incubation stage is shown in **Figure 2f,** 10 min after opening the In shutter. Few NWs are seen to nucleate in this short time. The top and side view SEM images of craters, in **Figures 2g** and **2h** respectively, show the representative well developed cratered surface morphology obtained after a prolonged growth. The top view SEM image of a growth surface in **Figure 2i** shows NWs emerging form the craters along two opposite <110> directions. Remarkably, we find that ripening the Au droplets in a rich As environment completely passivates the surface and prevents the nucleation of craters and as a consequence, prevents the formation of NWs (**Figure 2j**).

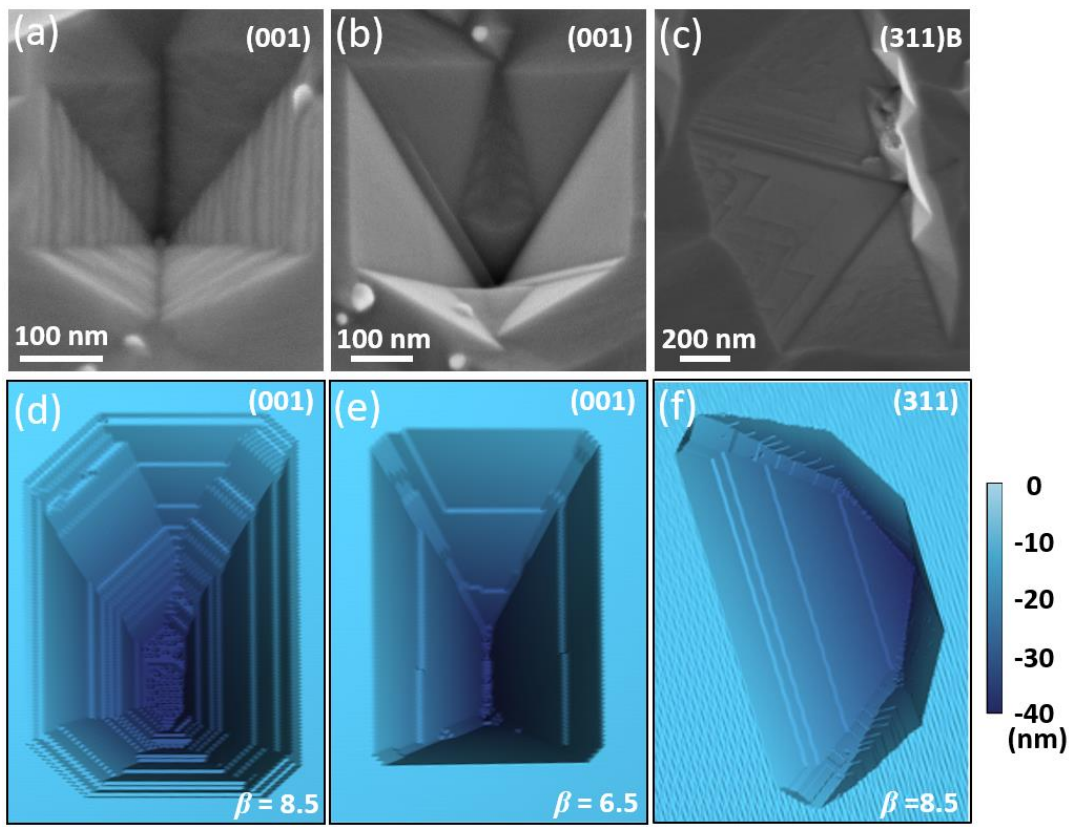

**Figure 3.** SEM images of a crater formed on a (001) surface at **(a)** low and **(b)** high temperature, with stepped and smooth facets, respectively; **(c)** SEM top view of a crater on a (311)B surface. **(d-f)** Craters obtained by 3D-Monte Carlo simulations: **(d-e)** dependence of the density of steps on the craters side facets on temperature ($T$), parameter $\beta \sim 1/T$. **(f)** typical crater on a (311) surface.

The formation of craters on the substrate surface was simulated using the kinetic Monte Carlo procedure. The model takes into account three dynamical processes: adsorption of the InAs particles at the sample surface, diffusion on the surface, evaporation and detachment of the particles from different surface sites. We start with a flat (001) or (311) surface with N = 350 × 350 sites and with a small hole at the central position, which we assume is created by the gold droplet (for more details see supplementary information). As seen in the SEM images in **Figures 3a-c**, the facets of the real crater might develop surface steps. The simulated craters show striking resemblance to the craters that form experimentally. The results of the simulations for two different values of the parameter $\boldsymbol{\beta}$ = $1/(k_B T)$, i.e., $\boldsymbol{\beta}$ = 8.5 and $\boldsymbol{\beta}$ = 6.5, for the craters on (001) surface are presented in **Figures 3d** and **3e,** respectively. **Figure 3f** shows the results of simulations for the crater formed at the higher temperature on a (311) surface. We note, that in both, **Figure 3c** and **3f**, one of the {111} facets of the crater is much larger than the other, as predicted from the tilt between the (311) and (001) surfaces. As can be seen in the **Figures 3d-f**, the calculated {111} side walls of the craters consist also of segments separated by small (001) steps. The smaller is $\boldsymbol{\beta}$, i.e., the higher is the temperature, the longer are the {111} segments and smoother the side walls. Their elongated shape was recovered by taking into account in the values of the next nearest neighbour interaction constants and occupation numbers the anisotropy present at the (001) surface. This anisotropy, which favours the {111} side facets, is important also for further growth of the NWs in the crater.[32] The calculations showed that the asymmetrical shape of the craters on the (311) surface is determined by the values of the detach energies. Also the shape of the craters on (001) surface, i.e., the proportion between various {111} and {110} side walls as well as the shape and size of the bottom of the crater, depends primarily on the energy needed to detach a particle from the bottom of the crater, as shown in the Supplementary Information (**Figure S7**). We found that elevated temperature may also promote the growth of the (001) base of the crater by modifying the relative surface diffusion length over the (001), (111)B and (110) surfaces (**Figure S3**).

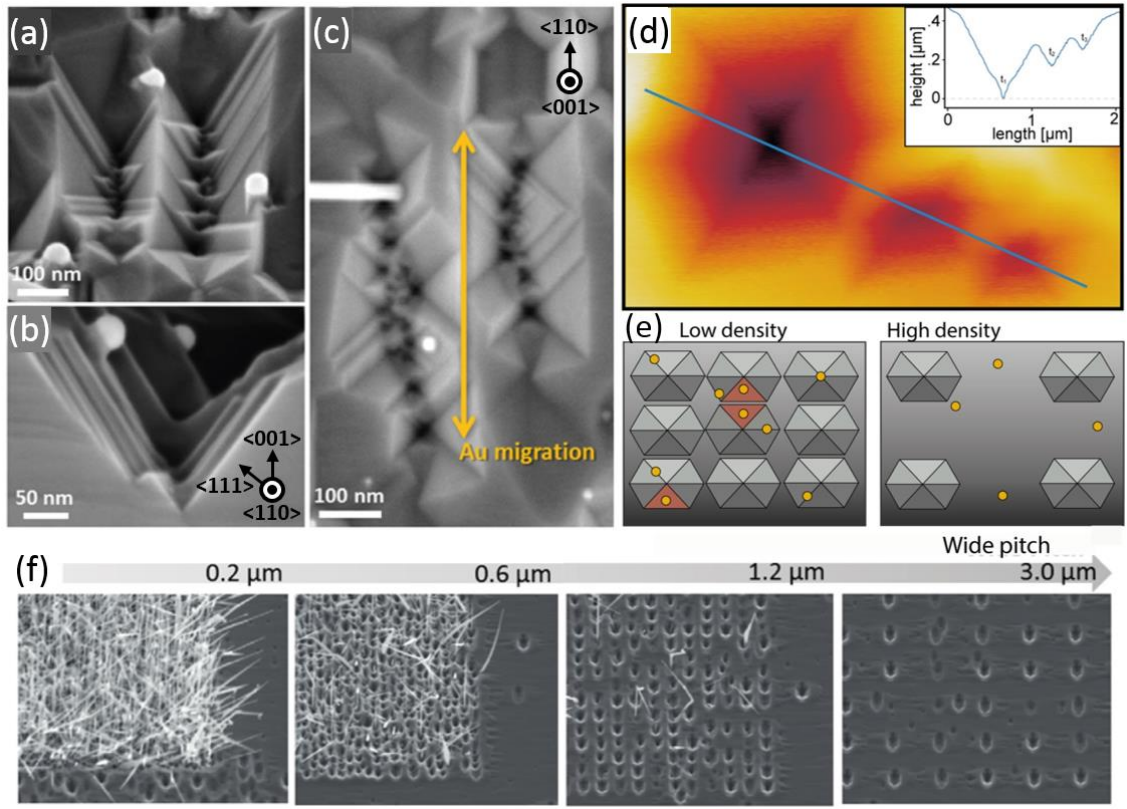

**Figure 4.** SEM images showing **(a)** 45° bird-eyes, **(b)** side, and **(c)** top views of a trail of craters created by Au migration along the <110> direction. **(d)** AFM image of such a trail with three sequential craters and a depth profile given in the inset. **(e)** Illustration of Au droplet migration and {111}B facet density generated by nucleation of craters for low and high density of Au droplets. The optimal configuration for NW nucleation is highlighted by red facets. **(f)** SEM images showing the dependence of the NWs' density on the density of Au droplets, for four samples patterned with different pitch.

As for the dynamics of the Au droplets, trails of craters formed by the diffusing Au droplets are clearly observed in **Figures 4a-c**. We found that those trails always stretch along the <110> direction, which is thus the easy axis for the Au droplet diffusion. An AFM image of such a trail with three sequential craters can be seen in **Figure 4d**, along with a depth profile, which is given in the inset. The discreteness of this trail may suggest that the Au droplet at an early stage diffuses out of the crater and settles at its (001) rim, where it nucleates a neighbouring crater.

To further study the dynamics of the Au droplets we have investigated Au patterned (001) substrates under high In flux conditions. We have patterned Au pads with varying densities, as shown schematically in **Figure 4e**. These pads are annealed to form droplets that nucleate nano-craters at their designated locations. The Au droplets diffuse out of the initial crater they nucleated. At that stage, the growth conditions do not support nucleation of additional craters. Therefore, in the most dilute case the Au droplets do not get captured in a crater and NWs do not form (see **Figure 4f**). As shown in **Figure S3**, the Au droplets were also found to be immobile on the large (001) crater bottoms of the high temperature patterned growth. The probability for the Au droplet to be recaptured in a crater grows with growing density of craters and, therefore, also with the total area of {111}B surface within the Au droplet diffusion length. Correspondingly, the density of NWs increases. This clearly shows the role of the Au droplet in nucleating the crater through the full correlation between the position of the Au pads and craters that form as well as the mobility of the Au droplet during the crater incubation stage. We stress that the dual role of catalyst droplets makes positioned growth on (001) surfaces quite challenging in terms of tuning the growth conditions.

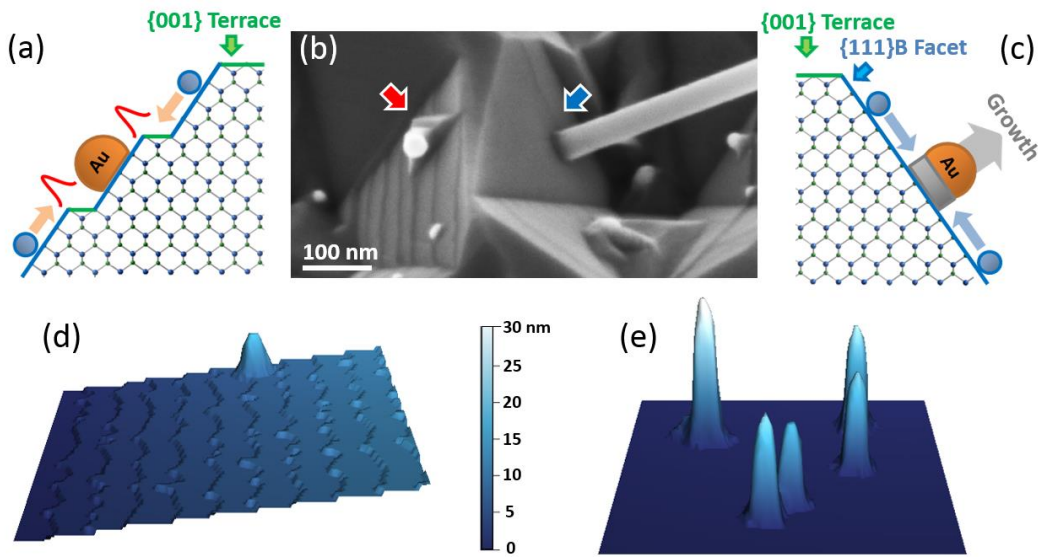

**Figure 5. (a)** and **(c)** Schematic illustrations of a Au droplet on a stepped and a smooth {111}B facet, respectively. **(b)** Corresponding SEM image of two such adjacent {111}B facets: the left hand one having a stepped surface and a gold droplet (marked in red) while the right hand shows a smooth facet with an emerging NW (marked in blue) of similar diameter to the droplet on the stepped surface. **(d)** and **(e)** The respective 3D-Monte Carlo simulations of a stepped and smooth {111}B surface.

As the highly diffusive side facets of the craters grow in size, so does the In collection area of Au droplets that happen to reside on them. This, along with ramping down the temperature, increases the In content in the droplets. At a certain stage the In collection area surpasses a critical dimension beyond which In supersaturation is achieved. The critical facet size will depend on the various growth parameters, for instance temperature, incoming fluxes, and droplet size.[33, 34] However, here enters yet another limiting factor, which is the morphology of the craters that can impede the surface diffusion of the In atom. This is strikingly captured by the SEM image in **Figure 5b**. It shows two back to back {111}B side facets of similar size, shape and local growth conditions that differ from one another merely by the occurrence or absence of surface steps. While the droplet on the flat facet catalyzed a prominent NW growth, the one on the stepped facet did not. We note that the different outcome did not originate from the two droplets being of different size, since the diameter of the droplet imaged is comparable to the diameter of the emerging NW imaged on the right hand side (which in turn is set by the diameter of the droplet that catalyzed its growth). This is commonly observed over the cratered growth surface.

This phenomenon can be associated with the high affinity of the surface atoms for bonding to the edges of steps present on the surface, as well as by finite probability to reflect from them, as sketched in **Figures 5a** and **5c**. As a result the local diffusion length will be truncated by the stepped boundaries on the {111}B side facets. A minimal collection area free of such steps is needed to support supersaturation conditions in the Au droplet that can initiate NW growth. This conclusion is confirmed by Monte Carlo simulations, in which a fast diffusion up on a step and a Schwoebel barrier (which prevents the particle from jumping back from the step) were added in order to simulate the NW growth.[35] Two cases of the same area of {111}B surface were considered, were one is smooth and the other one stepped. All the other parameters were kept the same. The comparison of the calculated rates of emerging NWs at a smooth and stepped {111}B surface is presented in **Figures 5d** and **5e**, respectively. It shows that indeed, the steps inhibit NWs growth. We therefore conclude that since for the onset of NWs a critical effective area for adatom diffusion should be secured, the morphology of the substrate surface plays a critical role in the process of nucleation of NWs.

Once the craters are fully formed and wherever their nano-facets are sufficiently flat and sufficiently large, the common preconditions that apply to (111)B substrates will determine the emergence of a NW from a droplet. Namely, very small droplets will fail to compete with the 2D growth of their surroundings. The large ones will not reach supersaturation and thus remain on the surface. A cross-section FE-SEM illustration, and HR-TEM images in **Figure S4** confirm that the structure of just nucleated NW is ZB, which is determined by the crystal structure of the (111)B side facet and the low In supersaturation within the crater. Shortly after emerging from the crater the structure of the NW switches from ZB to WZ as expected at the higher supersaturation outside of the craters and the given growth direction of <111>. This transition is seen for all NWs that terminate at the cleavage plane.

Intriguingly, some droplets remain trapped at the very bottom of the craters where the (001) surface dictates their growth in the <001> direction, naturally having the ZB structure all along (**Figure S5**). We speculate that the {111}B nano-facets, on either side of such NW, feed the droplet with In and, therefore, facilitate NW growth from the very bottom of the crater. Other droplets diffuse out of the craters and end up on the ridges where they cannot develop into NWs, due to the strong competition with the bulk growth. The same is true for droplets residing on the {011} facets of the craters. We find that within the cratered structure the only facets which allow collection of sufficient material into the Au droplets, thereby allowing them to nucleate NWs, are the two opposite {111}B facets present in each crater.

In conclusion, the growth of inclined NWs on a (001) or (311)B InAs substrate is shown to be governed by gold in its dual role. The gold is responsible first for the formation of a proper platform, from which the NWs can emerge, and then for guiding the growth of individual NWs. We show that the Au-assisted VLS growth of inclined NWs depends first and foremost on the Au-assisted formation and properties of craters with {111}B side facets, without which no NWs can emerge. More specially, the microstructure of the {111}B nano-facet within the craters, which are formed on the {001}-type surfaces, is critical to the growth of inclined NWs. While the craters formed on (001) surface have a regular hexagonal shape with different {111} nano-facets, in the case of the (311)B surface only one type of {111} nano-facet is large enough to allow the growth of NWs. Thus, in the latter case all the wires grow parallel and only (001) substrates are suitable for the formation of NWs intersections and networks. These observations - the formation of the craters and their structure as well as the dependence of the onset of NWs on the surface morphology of the {111}B nano-facets within the craters - were confirmed by Monte Carlo simulations.

The vision of this work seems to extend well beyond a specific III-V material or a particular crystal plane. Au induced formation of faceted craters on the (001) and its related vicinal surfaces, turns out to be a crucial process that takes place prior to Au-assisted growth of NWs from such surfaces. The (001) surface is the more common crystal plane which is compatible with microelectronics processes certainly in comparison to the (111) surface, more commonly used for NWs growth. In general, growth of inclining NWs opens possibilities such as merging on the one hand and more facilitated in-situ coating on the other hand.

**Supporting Information**

**Figure S1** shows SEM images of GaAs NWs samples grown on a (001) and (311)B GaAs surface. **Figure S2** is an AFM mapping and line scan profile of a crater having 6 different facets (hexagonal shape). **Figure S3** shows SEM images of a high temperature growth with a "flat bottom" crater (inset), an empty crater with a large (001) flat bottom, and the respective notations of facets surrounding such a crater as well as schematic illustration of the crater cross-section at 2 different temperatures. **Figure S4** shows a side view-SEM image of InAs NWs emerging from craters on a (001) surface with zoom-in onto the gold droplet emerging form a {111}B nano-facet, along with a respective schematic illustration as well as HR-TEM image showing the phase transition from ZB to WZ at the base of an InAs NW. **Figure S5** shows an occasional NW growing in the <001> direction among the ones growing <111> direction including zoom-in into its base and an SEM image showing its square cross-section. **Figure S6** is related to our Monte Carlo simulation and illustrates 6 parameters to rule the surface kinetics for growth. **Figure S7** is simulation data of two craters created on a (001) surface with different (001) bottom.

**Corresponding Author**

Dr. Hadas Shtrikman, Department of Condensed Matter Physics, Braun Center for Submicron Research, Weizmann Institute of Science, Rehovot 76100, Israel.

E-mail: hadas.shtrikman@weizmann.ac.il, Tel: +972-8-934-2585, Fax: +972-8-934-4128

## ACKNOWLEDGMENTS

First and foremost, we deeply thank Michael Fourmansky for his unstinting, professional technical assistance. All authors acknowledge partial financial supports of the Israeli Science Foundation (Grant No. 532/12 and Grant No. 3-6799), Israeli Ministry of Science (Grant No. 0321-4801 (16097)), BSF grant No. 2014098, and the Polish National Science Center (grant No. 2013/11/B/ST3/03934). Hadas Shtrikman, incumbent of the Henry and Gertrude F. Rothschild Research Fellow Chair. Microsoft Station Q is acknowledged for financial support. The authors are all grateful to Moty Heiblum for making this research possible.

# Supplementary Information

# Au-assisted substrate-faceting for inclined nanowire growth

*Jung-Hyun Kang[†*], Filip Krizek[‡*], Magdalena Zaluska-Kotur[§], Peter Krogstrup[‡], Perla Kacman[§], Haim Beidenkopf[†] and Hadas Shtrikman[†]*

[†] Dept. of Condensed Matter Physics, Braun Center for Submicron Research, Weizmann Institute of Science, Rehovot 76100, Israel
[‡] Center for Quantum Devices and Station Q Copenhagen, Niels Bohr Institute, University of Copenhagen, 2100 Copenhagen, Denmark
[§] Institute of Physics Polish Academy of Science, Al. Lotnikow 32/46, 02-668 Warsaw, Poland

The vision of this work seems to extend well beyond a specific III-V material or a particular crystal plane. Au induced formation of faceted craters on the (001) and its related vicinal surfaces, turns out to be a crucial process that takes place prior to gold assisted growth of NWs from such surfaces. The (001) surface is the more common crystal plane which is compatible with microelectronics processes certainly in comparison to the (111) surface, more commonly used for NWs growth. In general, growth of inclining NWs opens possibilities such as merging on the one hand and more facilitated in-situ coating on the other hand. In this SI we include some of the experimental data we were unable to incorporate in our main text, in particular the validity of our observations to the (001) and (311)B growth surfaces of GaAs. In addition we show AFM of a single crater, trapping of the gold droplet at the flat (001) bottom of a crater, the transition from ZB to WZ at the base of the NW and the occasional growth of a square, ZB NW in the <001> direction. Description of the Monte Carlo process used along with a couple of additional results is given at the end of the section.

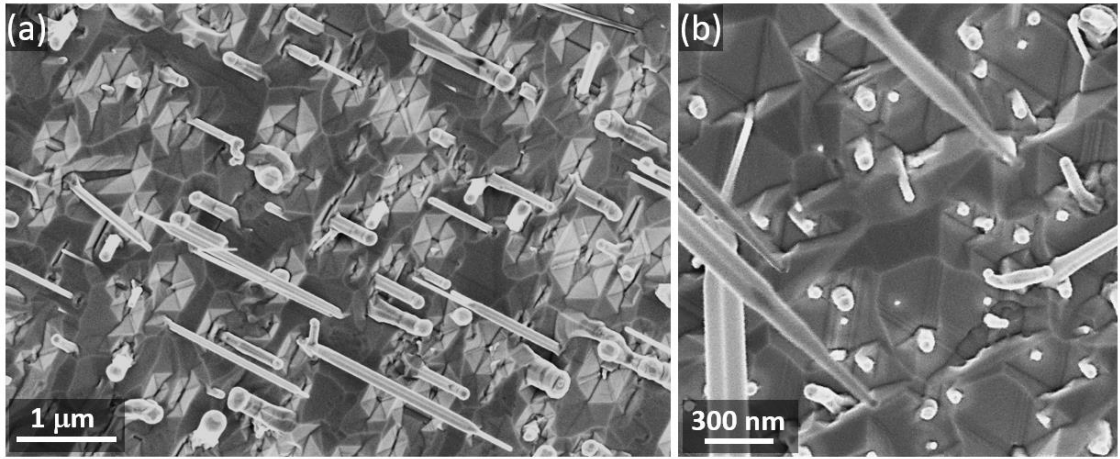

**Figure S1**: Top view SEM images of GaAs NWs sample grown on **(a)** a (001) and **(b)** (311)B GaAs surface. As in the case of InAs, on the GaAs (001)-type surfaces craters form with are clearly seen. Also here the 25° tilt of the craters on the (311)B surface diminishes one set of {111} facets leaving only one set of {111} facets, which is observed in the SEM image.

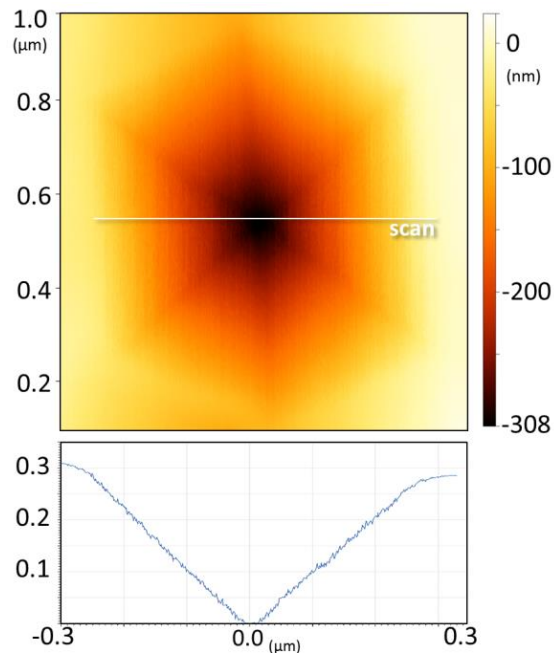

**Figure S2.** AFM mapping and line scan profile of the crater having 6 different facets (hexagonal shape).

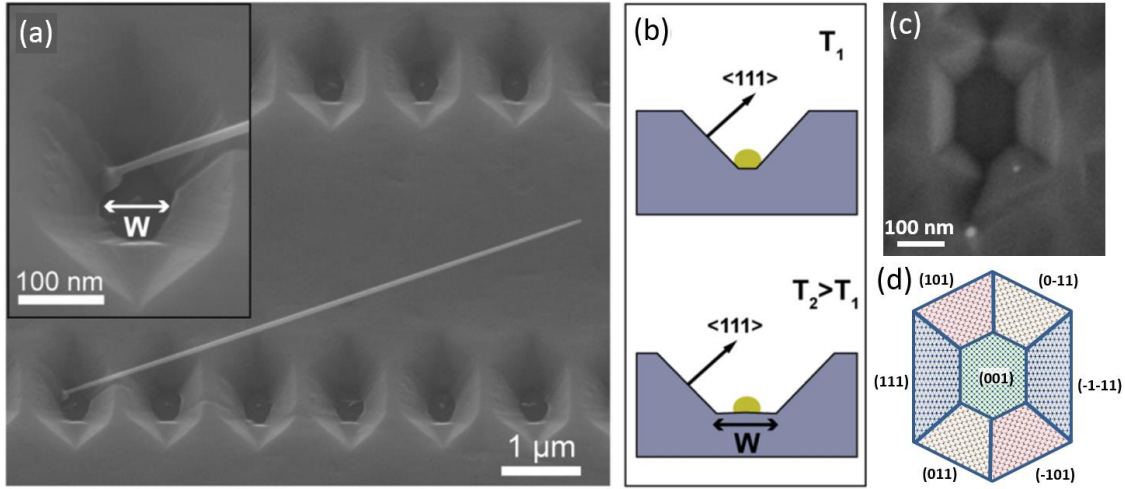

**Figure S3. (a)** SEM images showing a high temperature growth (~510 °C), which results in an increase of the (001) area at the bottom of the craters. The diffusion of the Au droplets is highly suppressed over the exposed (001) surface and thus they mostly remain at the crater center. The inset SEM image shows such "flat bottom" crater where an inclined NW emerges from the edge of the {111}B nano-facet. W is the distance between the edges of both {111}B facets within the crater. **(b)** Illustration of the crater cross-section at temperatures T1 and T2, where T2 is higher than T1; we expect the {111}B facet dimensions to scale as W~1/T. **(c)** an SEM image of an empty crater with a large (001) flat bottom. **(d)** The respective notations of facets surrounding such a crater.

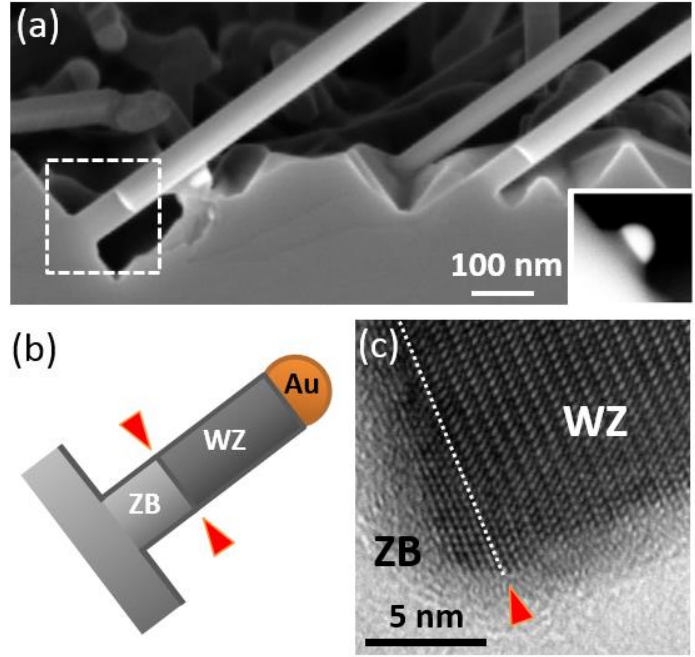

**Figure S4. (a)** Side view -SEM image of InAs NWs emerging from craters on a (001) surface showing the transition from ZB to WZ, as illustrated in **(b)**; inset: side view of a gold droplet emerging form a {111}B nano-facet. **(c)** HR-TEM image showing the phase transition from ZB to WZ at the base of an InAs NW.

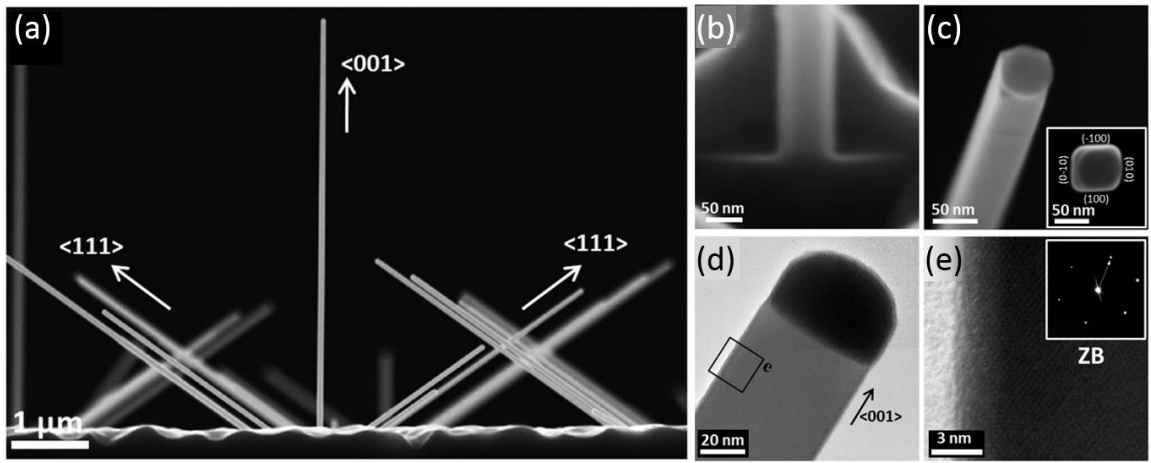

**Figure S5. (a)** FE-SEM side-view, **(b-c)** enlarged-SEM, **(d)** TEM, **(e)** HR-TEM, and SADF pattern (inset) images showing vertical InAs NWs with ZB structure grown on (001) substrate.

**Monte Carlo Simulation**

The simulated system consists of atoms at a crystal lattice of ZB symmetry and a layer of free, diffusing adatoms. In the process of adsorption or desorption the atoms exchange between these two subsystems. Several phenomena, which take place in the microscale, like desorption of In atoms, adatom diffusion and incorporation of adatoms at steps and kinks, have an impact on the surface morphology during the growth. Individual probabilities for the occurrence of each of the processes are set. We assume that whenever an adatom appears at a step or a kink, it is immediately incorporated. Taking into account that the growth happens in the excess of As atoms, we assume that only In atoms execute all surface processes with given rates, whereas As builds into the lattice completing the proper crystal structure.

Adatom desorption and diffusion depend on the energy $E_i$ of adatom at a given site j and on the additional energy barrier $E_B$. Both processes are activated by temperature. The shape of the crystal surface is controlled by time evolution and the time scale is given by the diffusion rate. The craters with <111>-oriented walls are formed on the (001)-type surface due to 1) evaporation, which is active in the central positions, where the gold influence is assumed and 2) detaching from surface steps. As illustrated in **Figure S6**, we denote by $\boldsymbol{R_t}$ , $\boldsymbol{R_s}$, and $\boldsymbol{R_b}$ the rates of detach from different surface steps: top (t), side (s) and bottom (b). $\boldsymbol{D}$ and $\boldsymbol{P_{ev}}$ are the diffusion and evaporation probabilities, respectively. Each probability depends on temperature T and on the difference between the energy of the site and the barrier:

$$\boldsymbol{R}, \boldsymbol{P}, \boldsymbol{D} = \vartheta e^{-\Delta E / k_B T}, \qquad (1)$$

where

$$\Delta E = E_B + J_1 n_1 + J_2^A n_2^A + J_2^B n_2^B + J_3 n_3. \qquad (2)$$

denotes the energy needed to detach a particle from a given site, which is a sum of the bonding energy and the barrier. The energy barrier $E_B$ depends on the position of the step: in the simulations different values of $E_B^b$, i.e., the barrier at the bottom of the crater, were taken to change the shape of the crater. The barrier at the side wall of the crater $E_B^S = 0$. Due to the anisotropy, which is present on the InAs substrate surface, the barrier $E_B^t$ at the kink at the top of the crater depends on the step direction and has different values $E_B^t = 0.2, 0.2, 0.4$ along the [110], [1-10] or [100] axis. At the bottom of crater the desorption energy was decreased by $E_B^b = -3.6$. Particles at the top edge of the crater were desorbed with the same rate for all temperatures. Diffusion $D$ of particles was assumed to be anisotropic and the fastest along [110] direction.

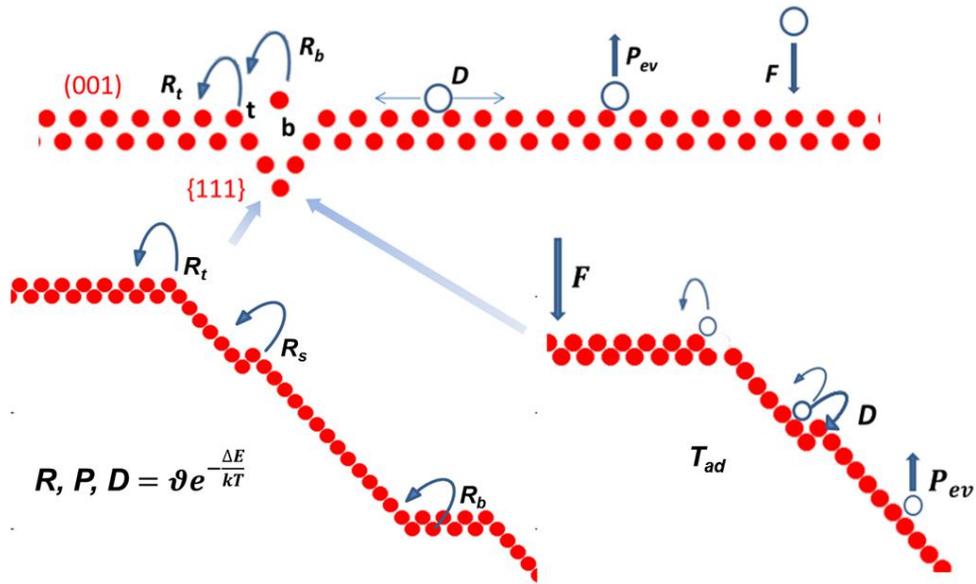

**Figure S6.** Illustration showing 6 parameters to rule the kinetics: $R_t$, $R_s$ and $R_b$ are rates of detach from surface steps $t$, $s$ and $b$, $P_{ev}$ - evaporation from surface is active for site positions where gold influence is assumed while the external flux of particles $F$ is active outside this region; diffusion $D$ and the adsorption at a kink $T_{ad}$ are present everywhere.

The bonding energy is given by the interaction constants $J_i$ multiplied by the appropriate number of occupied nearest neighbour ($n_1$), next nearest ($n_2$) and third nearest ($n_3$) sites and denotes what energy is needed to detach a particle from the surface. It is taken that all particles attach at kink positions with the same probability equal to one.

Due to the anisotropy, two values for the next nearest neighbour interaction constant $J_2$ and occupation number $n_2$ are assumed and denoted A and B. The following parameters were used in the simulations: $J_1 = 1, J_2^A = 0.6, J_2^B = 0.65, J_3 = 0.6$. The illustration of all parameters used in our kinetic Monte Carlo simulations is presented in **Figure S6**.

The shape of the craters, i.e., the proportion between various {111} and {110} side walls as well as the shape and size of the bottom of the crater can be changed by changing the parameter $E_B^b$, as shown in **Figure S7** for the craters on a (001) surface.

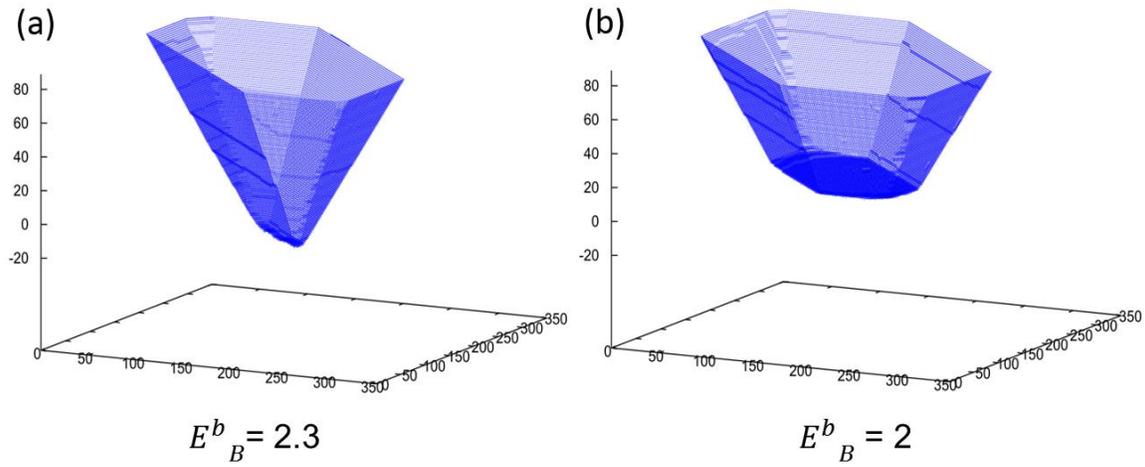

**Figure S7.** The two craters created on a (001) surface with different size of the (001) bottom obtained for two different values, -2.3 and -2, of the parameter $E_B^b$ with $\beta = 1/(k_BT) = 7$ in both cases.